\documentclass[aps,prl,twocolumn,a4paper,superscriptaddress]{revtex4-1}
\usepackage{amssymb}
\usepackage[utf8x]{inputenc}
\usepackage[]{graphicx}
\usepackage{bm,color,subfigure,amsmath}
\begin{document}
\title{Generation linewidth of mode-hopping spin torque oscillators}

\author{Ezio Iacocca}
\email{ezio.iacocca@physics.gu.se}
\affiliation{Physics Department, University of Gothenburg, 412 96, Gothenburg, Sweden}

\author{Olle Heinonen}
\affiliation{Materials Science Division, Argonne National Laboratory, Lemont, Illinois 60439, USA}

\author{P. K. Muduli}
\affiliation{Physics Department, University of Gothenburg, 412 96, Gothenburg, Sweden}
\affiliation{Department of Physics, Indian Institute of Technology Delhi, Hauz Khas 110016, New Delhi, India}

\author{Johan \AA{}kerman}
\affiliation{Physics Department, University of Gothenburg, 412 96, Gothenburg, Sweden}
\affiliation{Material Physics, School of ICT, Royal Institute of Technology, Electrum 229, 164 40, Kista, Sweden}

\begin{abstract}
Experiments on spin torque oscillators commonly observe multi-mode signals. Recent theoretical works have ascribed the multi-mode signal generation to coupling between energy-separated spin wave modes. Here, we analyze in detail the dynamics generated by such mode coupling. We show analytically that the mode-hopping dynamics broaden the generation linewidth and makes it generally well described by a Voigt lineshape. Furthermore, we show that the mode-hopping contribution to the linewidth can dominate in which case it provides a direct measure of the mode-hopping rate. Due to the thermal drive of mode-hopping events, the mode-hopping rate also provides information on the energy barrier separating modes and temperature-dependent linewidth broadening. Our results are in good agreement with experiments, revealing the physical mechanism behind the linewidth broadening in multi-mode spin torque oscillators.
\end{abstract}
\maketitle

Nanoscopic excitation of high-amplitude magnetization dynamics has recently emerged due to the discovery of the spin-transfer torque (STT) effect and advances in nano-fabrication~\cite{Slonczewski1996,Berger1996}. STT describes the momentum transfer from spin-polarized electrons to a local magnetization and therefore provides a direct coupling between dc charge currents and magnetization dynamics. Depending on external conditions, a rich variety of physical phenomena with technologically interesting outcomes are possible, including different modes of spin wave generation~\cite{Slonczewski1999,Slavin2005,Bonetti2010,Madami2011,Bonetti2012,Dumas2013}, vortex gyration~\cite{Pribiag2007,Devolder2010,Dussaux2010,Petit2012}, and the nucleation and manipulation of magnetic droplet solitons~\cite{Hoefer2010,Hoefer2012,Mohseni2013,Mohseni2013b}. Regardless of the particular magnetization dynamics, devices where a stable oscillatory state can be achieved are generally referred to as spin torque oscillators~\cite{Silva2008,Ralph2008} (STOs), and are typically composed of two ferromagnetic layers decoupled by a non-magnetic spacer (although recent studies also report on STOs based on single ferromagnet layers~\cite{Sani2013}). STOs are engineered to enforce magnetization dynamics in one of the ferromagnetic layers (the ``free'' layer) whereas the second layer (the ``fixed'' layer) acts both as a polarizer and a reference to probe the dynamics via magnetoresistive effects~\cite{Baibich1988,Binasch1989,Tsoi1998,Julliere1975,Yuasa2004,Houssameddine2008}.

STOs have been traditionally regarded as single mode oscillators~\cite{Rezende2005,Slavin2009} based on the mode selection imposed by the balance of STT and magnetic damping as well as the survival of the mode with the lowest threshold. However, recent experiments have shown multi-mode generation in a large variety of geometries~\cite{Kiselev2004,Sankey2005,Zeng2010}, revealing evidence of mode-hopping~\cite{Krivorotov2008,Muduli2012,Muduli2012b}, periodic mode transitions~\cite{Bonetti2010,Bonetti2012}, and even coexistence~\cite{Dumas2013}. Furthermore, such a multi-mode generation leads to broader linewidths ascribed to the reduction of the magnetization dynamics coherence. In order to understand the underlying physics of these observations, a multi-modal theoretical description is required.

A first step towards this goal was recently proposed~\cite{Muduli2012,Muduli2012b,Heinonen2013} by extending the Slavin - Tiberkevich auto-oscillator theory~\cite{Slavin2009} for two coupled modes. The general form of such an extension was found to agree with the equations describing multi-mode ring lasers~\cite{Beri2008,Sande2008} and thus support mode-hopping. However, the model equations remained qualitative and their relation to experimental observables was not explored. Here, we investigate multi-mode STOs with a goal to {\em quantitatively} describe their generation linewidth and thus provide a direct connection with experimental quantities, revealing the underlying physical mechanism driving the dynamics.

A central result of this paper is the derivation of the expected linewidth of a two-mode oscillator in a mode-hopping regime. We will show that the linewidth is enhanced by multi-mode generation and is described by a Voigt lineshape. Mode-hopping events can be described by a Poisson process providing a purely Lorenztian contribution to the linewidth which dominates at high mode-hopping rates. Such rates are well described by an Arrhenius distribution, providing information on the energy barrier between the modes and furthermore explains temperature-driven linewidth broadening. The presented results offer means to experimentally access previously unexplored features of STOs and directly connect experiments with parameters in the theory.

The multi-mode model equations introduced in Ref.~\onlinecite{Muduli2012} originate from first principle calculations by considering e.g., two stable modes coupled by scattering processes in a magnon bath~\cite{Heinonen2013b}. As a consequence, additional damping, torque, and coupling terms arise. Here, we further incorporate thermal fluctuations following the scheme of Ref.~\onlinecite{Slavin2009}, where $\tilde{f}=f^R+if^I$ is a Gaussian distributed perturbation with real and imaginary components and second moment given by $\langle\tilde{f}(t)\tilde{f}(t')\rangle=p_i\Delta\omega\delta(t-t')$, where $p_i$ is the power of the $i$-th mode and $\Delta\omega$ is the (linear) STO generation linewidth derived from the Slavin-Tiberkevich framework. Performing some algebra~\cite{SuppMat}, the coupled equations can be cast in terms of the variables $\theta$ and $\psi$ which map, respectively, the modes' energy and their phase difference onto a two-dimensional phase-space
\begin{subequations}\label{eq:thetapsi}
\begin{eqnarray}
\label{eq:theta}
    \dot{\theta} &=& \cos{\theta}\Gamma_Gp\frac{1-\sin{\theta}}{2\omega_1}\left[\bar{Q_1}-\bar{Q_0}+\xi(\bar{P_1}-\bar{P_0})\right]\nonumber\\   &&-\cos{\theta}\Gamma_Gp\frac{1+\sin{\theta}}{2\omega_2}\left[\bar{Q_2}-\bar{Q_0}+\xi(\bar{P_2}-\bar{P_0})\right]\nonumber\\
    &&+K(1-\sin{\theta})\sqrt{\frac{\omega_2}{\omega_1}}\cos{(\phi_c-\psi)}\nonumber\\
    &&-K(1+\sin{\theta})\sqrt{\frac{\omega_1}{\omega_2}}\cos{(\phi_c+\psi)}\nonumber\\    &&+\sqrt{\frac{2}{p}}\left[\cos{\frac{\theta}{2}}(f_2^R-f_1^R)-\sin{\frac{\theta}{2}}(f_2^R+f_1^R)\right],\\
\label{eq:psi}
    \dot{\psi} &=& \frac{pN_0}{2}\left[\frac{1-\sin{\theta}}{\omega_1}-\frac{1+\sin{\theta}}{\omega_2}\right]\nonumber\\
    &&+K\frac{1-\sin{\theta}}{\cos{\theta}}\sqrt{\frac{\omega_2}{\omega_1}}\sin{(\phi_c-\psi)}\nonumber\\
    &&-K\frac{1+\sin{\theta}}{\cos{\theta}}\sqrt{\frac{\omega_1}{\omega_2}}\sin{(\phi_c+\psi)}\nonumber\\    &&+\sqrt{\frac{2}{p}}\left[\frac{\cos{\frac{\theta}{2}}(f_2^I-f_1^I)-\sin{\frac{\theta}{2}}(f_2^I+f_1^I)}{\cos{\theta}}\right].
\end{eqnarray}
\end{subequations}

Here, the total power $p=p_1+p_2$ is enforced to be constant, given that $p_1=p\cos^2{\left(\theta/2+\pi/4\right)}$, $p_2=p\sin^2{\left(\theta/2+\pi/4\right)}$, and the condition $|\theta|\leq\pi/2$ is satisfied. The coefficients $\bar{Q_i}$ and $\bar{Q_0}$ are the diagonal and off-diagonal damping terms whereas $\bar{P_i}$ and $\bar{P_0}$ are the diagonal and off-diagonal STT terms. They are related to the auto-oscillator terms as $\Gamma_+(p_i)=\Gamma_G(1+\bar{Q_i}p_i+\bar{Q_0}p_j)$ and $\Gamma_-(p_i)=\xi\Gamma_G(1-\bar{P_i}p_i-\bar{P_0}p_j)$, where $i,j$ are indices for each of the two modes, $\xi=I_{dc}/I_{th}$ is the supercriticality, $I_{dc}$ is the bias current, $I_{th}$ is the threshold current for auto-oscillations, $\Gamma_G=\alpha\omega_o$, $\alpha$ is the Gilbert damping, and $\omega_o$ is the ferromagnetic resonance (FMR) frequency. A coupling term is included as a complex factor $Ke^{i\phi_c}$, with amplitude and phase $K$ and $\phi_c$, respectively. As we will see below, these terms determine the multi-mode dynamics and impact the generation linewidth.

\begin{figure}[t]
\centering\includegraphics[width=3.3in]{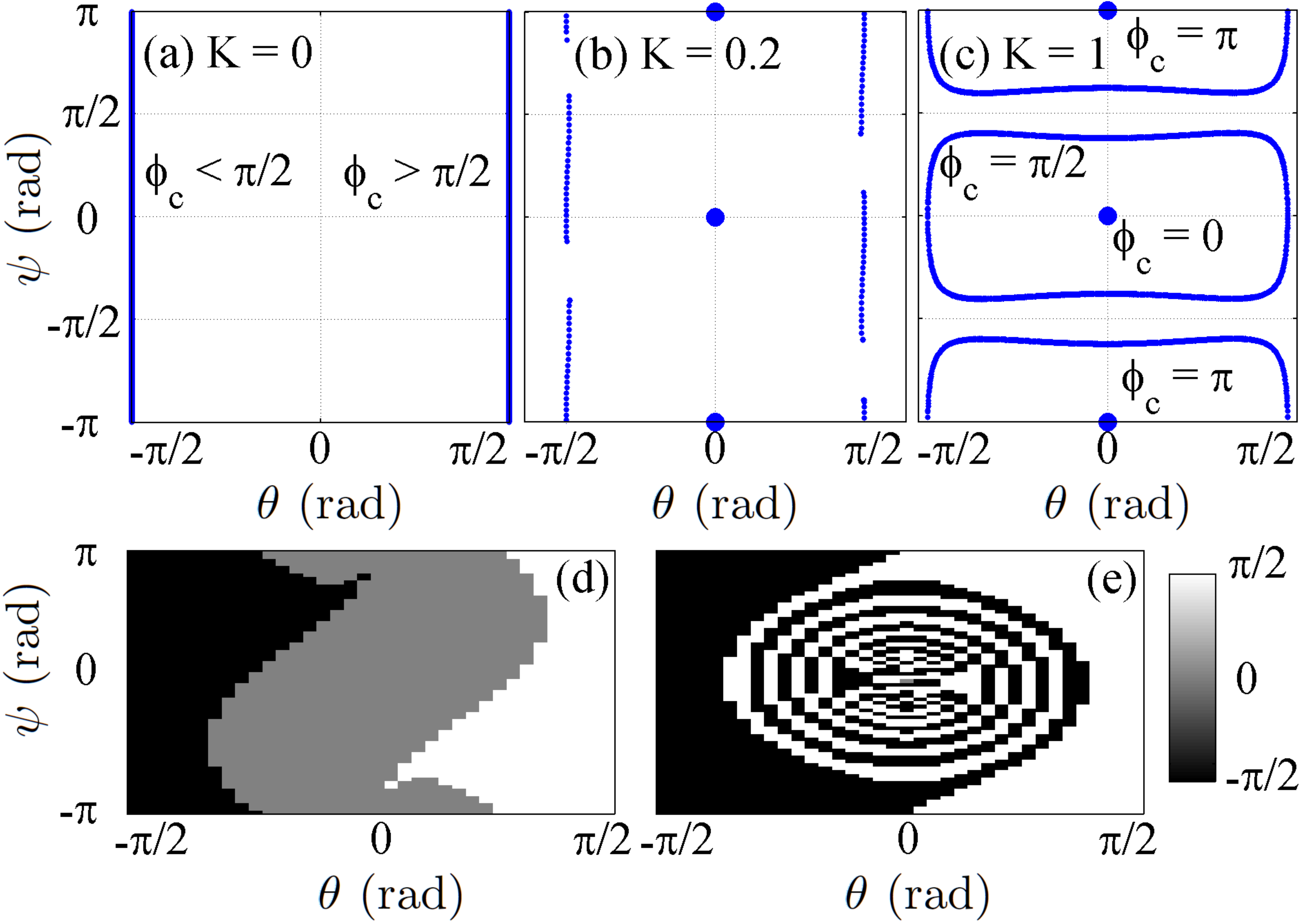}
\caption{ \label{fig1} Phase spaces spanned by $\theta$ and $\psi$ for (a) $K=0$: single mode; (b) $K=0.2$: near-single modes and coexistence; and (c) $K=1$: coexistence and periodic mode transitions. For the case $K=0.2$, the basins of attraction are shown for (d) $\phi_c=0$ and (e) $\phi_c =\pi/2$. In the latter, the spiral feature is reminiscent of a particle with friction in a double potential well. }
\end{figure}

Despite the algebraic complexity of Eq.~(\ref{eq:thetapsi}), two limiting cases are readily obtained when the thermal fluctuations are neglected~\cite{SuppMat}. A single mode exists when $\theta=\pm\pi/2$ and $K\rightarrow 0$ i.e., the coupling between modes is negligible [Fig.~\ref{fig1}(a)]. Note that $\psi$ diverges in this limit since a phase difference cannot be defined. From linear stability analysis~\cite{SuppMat} we find that the modes are independently stable if $\bar{Q_i}+\xi\bar{P_i}<\bar{Q_0}+\xi\bar{P_0}$. On the other hand, periodic mode transition and coexistence are possible when $K>0$. Linear stability analysis~\cite{SuppMat} demonstrates that each scenario depends on the coupling phase $\phi_c$, as indicated in Fig.~\ref{fig1}(c). Between the two limiting cases described above, near-single modes and coexistence are possible [Fig.~\ref{fig1}(b)]. The near-single mode scenario is of particular interest since each mode has a finite energy leading to thermally driven mode-hopping, as we will show below. Furthermore, the basin of attraction becomes strongly dependent on $\phi_c$, as shown in Fig.~\ref{fig1}(d-e) when $\phi_c=0$ and $\phi_c=\pi/2$. In the latter case, the spiral feature is reminiscent of a particle with friction in a double potential well~\cite{Ott1994} i.e., two stable modes separated by an energy barrier. In the following, we set $\phi_c=\pi/2$ in order to favor a mode-hopping scenario between two near-single modes.

\begin{figure}[t]
\centering\includegraphics[width=3.3in]{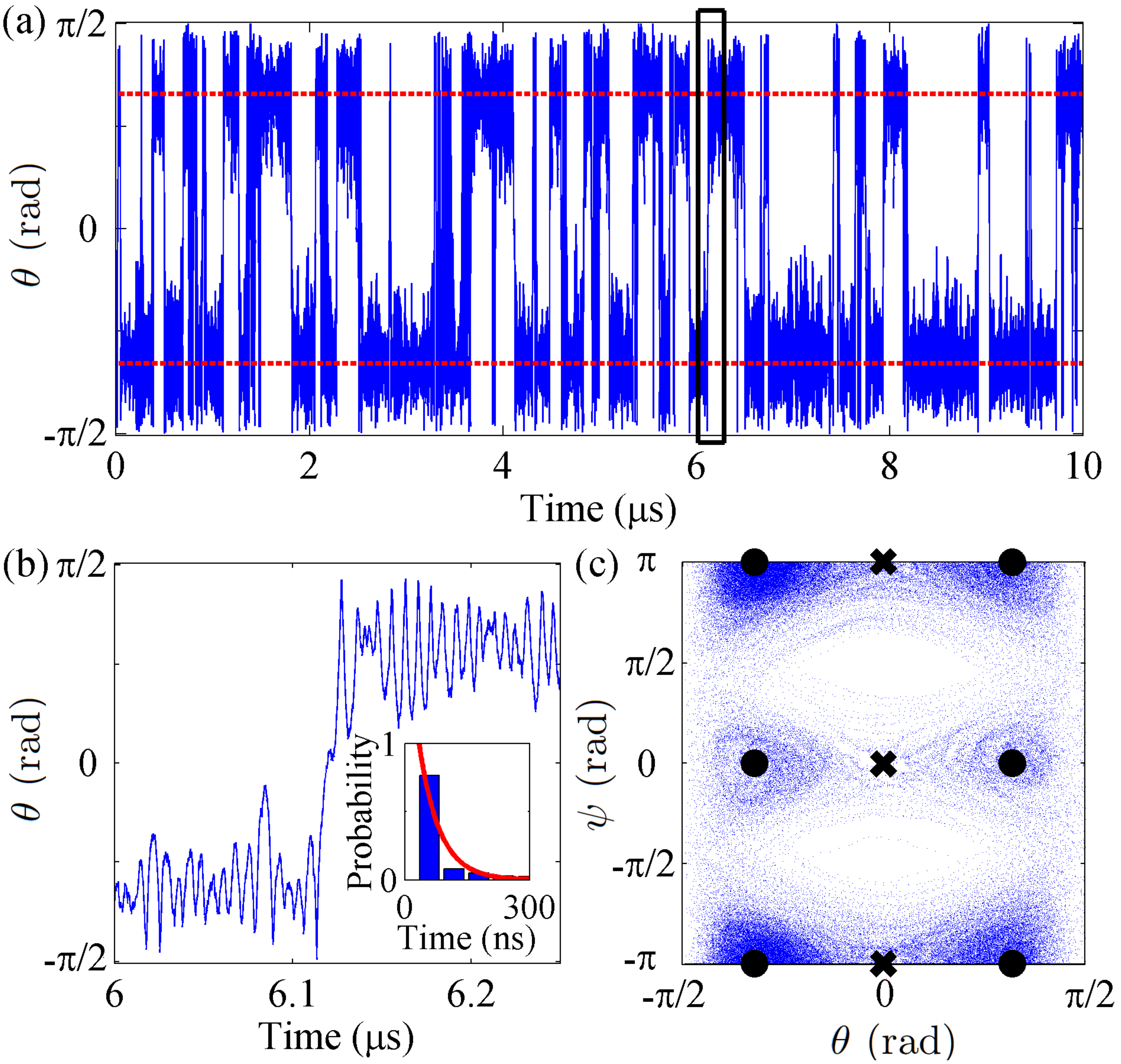}
\caption{ \label{fig2} (a) Time trace of $\theta$ exhibiting mode-hopping events between $\pm\langle\theta_o\rangle$ (red dashed lines). The section in the black box is detailed in (b), where the underlying relaxation frequency is observed. The inset shows the exponential distribution of the time difference between mode-hopping events, in agreement with a Poisson process. (c) Phase space of the time trace (a) showing that the stable modes (fixed points indicated by black dots) are thermally driven to mode-hop via saddle points (black crosses). }
\end{figure}

A typical time-trace exhibiting thermally driven mode-hopping is shown in Fig.~\ref{fig2}(a), where $K=0.3$ and we assume parameters consistent with the STO used in Ref.~\onlinecite{Muduli2012b}. Such a device consists of a $4.5$~nm thick Permalloy free layer with saturation magnetization $\mu_oM_S\approx0.88$~T, exchange length $\lambda_{ex}=5$~nm, and $\alpha=0.01$. An external field $\mu_oH_a=1$~T is applied at $80$~deg with respect to the Permalloy film plane. Whereas the current in [30] was confined to flow perpendicularly to the plane by patterning an elliptical 50~nm~$\times$~150~nm nanocontact, we here, for simplicity, assume a circular nanocontact of radius $R_c\approx40$~nm which has a similar effective current-carrying area with an assumed supercriticality $\xi\approx1.1$. In the two-mode oscillator framework, such parameters are mapped to $\omega_o/2\pi=11.94$~GHz, $p\approx0.017$, $\bar{Q}\approx4.6\omega$, $\bar{P}\approx\omega$, $\Gamma_G/2\pi\approx120$~MHz, $N_0/2\pi\omega\approx68$~GHz, $\omega/2\pi\approx13.13$~GHz, and $\Delta\omega/2\pi=0.6$~MHz. These parameters agree fairly well with the near-threshold generation of the real device. In order to complete the analytical description, we assume parameters for the off-diagonal terms $\bar{Q}_0=2\bar{Q}$ and $\bar{P}_0=2\bar{P}$, providing stability for both modes.

A particular mode-hopping event is indicated by the black box in Fig.~\ref{fig2}(a) and detailed in Fig.~\ref{fig2}(b). In this figure, the intrinsic relaxation frequency of the STO is also apparent, related to its strong nonlinear coefficient $N_0$~\cite{SuppMat}. On the other hand, the mechanism for the mode-hopping events can be clearly illustrated in the $(\theta,\psi)$ phase space shown in Fig.~\ref{fig2}(c). Here, the hopping between the different stable fixed points (indicated by black dots) occur via saddle points (indicated by black crosses). This picture also agrees with the abovementioned similarity between this system and a double potential well.

The generation linewidth of the resulting dynamics can be analytically estimated by means of the auto-correlation function of the two-mode oscillator defined as
\begin{equation}
\label{eq:autocorr}
    \mathcal{K} = \langle\left[c_1(t)+c_2(t)\right]\left[c_1^*(t')+c_2^*(t')\right]\rangle,
\end{equation}
where $c_i=\sqrt{p_i}e^{i\phi_i}$ is the complex amplitude of each mode, related to its power, $p_i$, and phase $\phi_i$. By manipulating the phases of each mode, it can be shown that the auto-correlation depends only on the second moment of the phase difference $\psi$~\cite{SuppMat}, thus providing a tractable expression for $\mathcal{K}$. Consequently, the problem is reduced to analyzing the thermally-induced behavior of $\psi$.

To proceed, two vastly different time-scales are identified: \emph{(i)} the perturbation introduced by thermal fluctuations, and \emph{(ii)} mode-hopping events. The former has a short time-scale and we will assume that the perturbation is small. The latter occurs as sharp phase-slips on a longer time-scale [Fig.~\ref{fig2}] and cannot be analytically obtained from Eq.~(\ref{eq:thetapsi}). Consequently, we will incorporate such events as an additional phase parametrized as a random train of pulses. In the following we derive the second moment contribution of each fluctuation source.

The perturbations of the phase difference can be estimated from the linearized coupled equations in phase and power -- as suggested in Ref.~\onlinecite{Slavin2009} -- by assuming a well-defined energy state for each mode, $\langle\theta_o\rangle$, as shown by red dashed lines in Fig.~\ref{fig2}(a). Performing a lengthy algebraic manipulation~\cite{SuppMat}, we can express the power and phase fluctuations as the coupled set of equations
\begin{subequations}\label{eq:deltappsi}
\begin{eqnarray}
\label{eq:deltap}
    \dot{\delta p} &=& C_{pp}\delta p + C_{p\psi}\delta\psi + f_p,\\
\label{eq:deltapsi}
    \dot{\delta\psi} &=& C_{\psi p}\delta p + C_{\psi\psi}\delta\psi + f_\psi,
\end{eqnarray}
\end{subequations}
where the coefficients are given in the Supplementary Material~\cite{SuppMat} and it is assumed that $p=p_o+\delta p$ and $\psi=\psi_o+\delta\psi$ satisfy the conditions $\delta p \ll p_o$ and $\delta\psi\ll\psi_o$.

In general, Eq.~(\ref{eq:deltappsi}) can be solved by the standard method of variation of parameters, as detailed in the Supplementary Material~\cite{SuppMat}. Such a solution leads to second moments proportional to exponential functions of $|\tau|=|t'-t|$. By performing a Taylor expansion to second order, the self- and cross-correlation second moments are
\begin{subequations}\label{eq:variance}
\begin{eqnarray}
\label{eq:selfterm}
    \langle\psi_{i}(t)\psi_{i}(t')\rangle &=& \gamma_{L,ii}|\tau|+\gamma_{G,ii}|\tau|^2,\\
\label{eq:crossterm}
    \frac{\langle\psi_{i}(t)\psi_{j}(t')\rangle}{\cos{\langle\theta_o\rangle}} &=& \gamma_{L,ij}|\tau|+\gamma_{G,ij}|\tau|^2,
\end{eqnarray}
\end{subequations}
where the coefficients $\gamma_L$ and $\gamma_G$ are mode dependent and are generally functions of the coefficients of Eq.~(\ref{eq:deltappsi})~\cite{SuppMat}.

On the other hand, mode-hopping events can be described by a series of sudden jumps in the phase difference, separated by random, long time intervals. This description is proper of a Poisson process~\cite{Silva2010} which is only described by its rate, $\lambda$. Indeed, the distribution of the relative time between mode-hopping events is numerically found to agree with an exponential distribution, as shown in the inset of Fig.~\ref{fig2}(b). Finally, it is known that the second moment of such a process is simply $\lambda$, so the phase difference is enhanced by a factor $-\lambda|\tau|$.

Including the two contributions described above into Eq.~(\ref{eq:autocorr}), we obtain the approximate yet insightful form of the auto-correlation
\begin{eqnarray}
\label{eq:autocorr_final}
    2\mathcal{K} &\propto& (1-\sin{\langle\theta_o\rangle})e^{-\gamma_{L,ii}|\tau|}e^{-\gamma_{G,ii}|\tau|^2}e^{-\lambda|\tau|}\nonumber\\
    &&+(1-\sin{\langle\theta_o\rangle})e^{-\gamma_{L,jj}|\tau|}e^{-\gamma_{G,jj}|\tau|^2}e^{-\lambda|\tau|}\nonumber\\
    &&+\cos{\langle\theta_o\rangle}(e^{-\gamma_{L,ij}|\tau|}e^{-\gamma_{G,ij}|\tau|^2})^{\cos{\langle\theta_o\rangle}}e^{-\lambda|\tau|}\nonumber\\
    &&+\cos{\langle\theta_o\rangle}(e^{-\gamma_{L,ji}|\tau|}e^{-\gamma_{G,ji}|\tau|^2})^{\cos{\langle\theta_o\rangle}}e^{-\lambda|\tau|}.
\end{eqnarray}

Equation~(\ref{eq:autocorr_final}) is a central result of this paper. Clearly, as the mode-hopping rate $\lambda$ increases, there will be a cross-over to the temporal decay of the correlation dominated by decoherence arising from mode-hopping. The resulting lineshape is obtained by the Fourier transform of the auto-correlation Eq.~(\ref{eq:autocorr_final}), from which the linewidth can be extracted. Each term of the right hand side has a similar form which, after performing the Fourier transform, leads to a Lorentzian lineshape with a linewidth given by $\gamma_{L,ij}+\lambda$, convoluted by a Gaussian lineshape with a linewidth given by $4\sqrt{\gamma_{G,ij}\mathrm{ln}2}$. Such a convolution is defined as a Voigt lineshape. Consequently, the general lineshape obtained from the Fourier transform of Eq.~(\ref{eq:autocorr_final}) is expected to be a sum of Voigt functions. Noteworthy, the mode-hopping rate $\lambda$ enhances the linewidth of the Lorentzian components, contributing to spectral broadening as observed experimentally~\cite{Muduli2012b}. On the other hand, the Gaussian contribution here arises from the response of Eq.~(\ref{eq:deltappsi}) which is found to relax to zero~\cite{SuppMat}, i.e., the auto-correlation is lost after a finite time, leading to statistically independent modes and thus uncorrelated mode-hopping events. This mechanism has a different physical origin than the Gaussian lineshape that arises from a high temperature limit~\cite{Slavin2009} or $1/f$ noise~\cite{Keller2010}.

\begin{figure}[t]
\centering\includegraphics[width=3.3in]{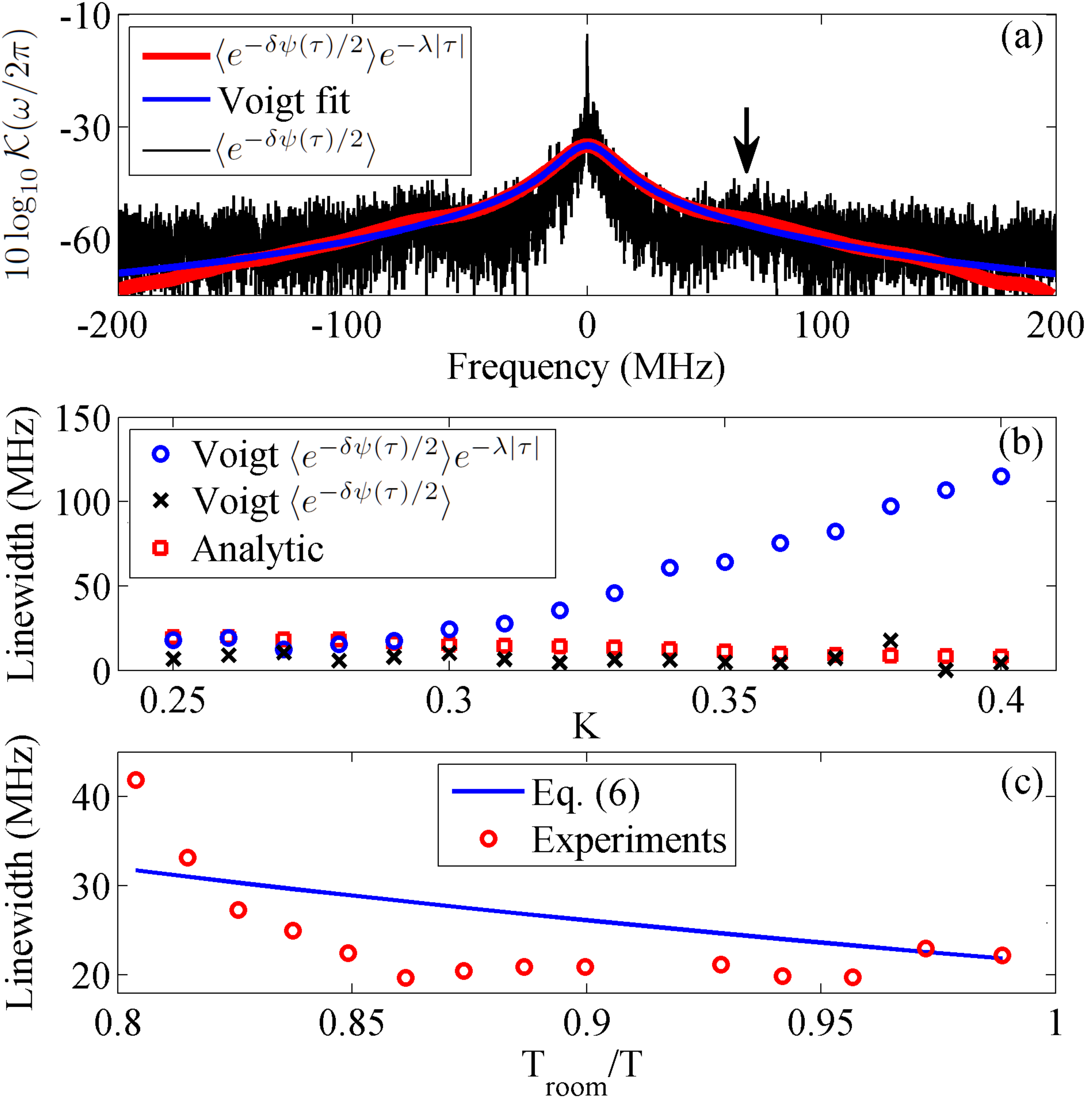}
\caption{ \label{fig3} (a) Fourier transform of the auto-correlation calculated from the numerical integration of Eq.~(\ref{eq:deltappsi}) (black). Mode-hooping events are included as a Poisson process with a mode-hopping rate calculated from the numerical solution of Eq.~(\ref{eq:thetapsi}) (red). The best Voigt fit is shown in blue. (b) Voigt linewidth as a function of the coupling strength $K$, obtained from the best fit of the auto-correlation (black) and including mode-hopping events (blue). The analytical estimate is shown by red squares. The mode-hopping events dominate the linewidth when $K>0.3$. (c) Expected temperature dependent linewidth (blue line) compared to the experimental linewidth~\cite{Muduli2012b} by assuming a constant $\Delta E$. }
\end{figure}

Numerically, the lineshape predicted from Eq.~(\ref{eq:autocorr_final}) can be obtained from the auto-correlation of $\delta\psi(t)$ multiplied by the Poisson factor with a mode-hopping rate $\lambda$ estimated from the time trace of Eq.~(\ref{eq:theta}). Such a lineshape is shown in Fig.~\ref{fig3}(a) by the red line for the parameters given earlier. We find the best Voigt fit following the approach of Ref.~\onlinecite{Liu2001}, as shown in the same figure by the blue line. For the chosen STO parameters, a single Voigt fit provides a good estimate of Eq.~(\ref{eq:autocorr_final}). The fitting procedure can be repeated as a function of $K$ from which we obtain the Voigt linewidths, $\Delta f$ (Half Width at Half Maximum), shown in Fig.~\ref{fig3}(b) by blue circles. These numerical results can be quantitatively compared with the analytical estimates from Eq.~(\ref{eq:autocorr_final}) and Eq.~(\ref{eq:variance}). For the chosen parameters, we obtain $\gamma_{G,ii}\rightarrow 0$ so that $\gamma_{L,ii}$ provides a good estimate for the linewidth, shown in Fig.~\ref{fig3}(b) by red squares. Clearly, the Voigt fit agrees well with the Lorentzian estimates when $K<0.3$ suggesting that the linewidth is otherwise dominated by mode-hopping, i.e., $\Delta f\rightarrow\lambda$. A second key result of this paper is that the obtained linewidth values \emph{quantitatively} agree with the reference experiment~\cite{Muduli2012} without any fitting parameters, but instead considering the mode-hopping as the physical mechanism behind linewidth broadening.

To further test the analytical estimates, we fit the spectrum of the phase difference auto-correlation shown by black lines in Fig.~\ref{fig3}(a). Multiple Voigt functions can be identified in this case, in agreement with Eq.~(\ref{eq:autocorr_final}). In particular, there is a narrow peak consistent with $\gamma_{L,21}\approx\gamma_{G,21}\rightarrow 0$. Noteworthy, sidebands corresponding to the oscillatory relaxation of the system are observed at about $\pm 80$~MHz (indicated by an arrow) which, together with the large fluctuations, allows us to  reliably fit only two Voigt functions. Independently of these difficulties, the wider Voigt linewidth [black marks in Fig.~\ref{fig3}(b)] is observed to follow the analytical trend, confirming that the linewidth enhancement is due to mode-hopping events.

The linewidth enhancement is consistent with the experimental observations close to a mode transition, indicating that the mode coupling increases in such a regime. Consequently, the linewidth provides a direct measure of $\lambda$. Assuming an Arrhenius distribution for the mode-hopping rate, it is thus possible to experimentally obtain information about the energy barrier, $\Delta E$, between two near-single modes e.g., as a function of the current,
\begin{equation}
\label{eq:Arrhenius}
    \Delta E(I_{dc}) = k_BT\log\frac{f_a}{\lambda},
\end{equation}
where $k_B$ is the Boltzmann constant, $T$ the temperature, and we assume that the attempt frequency $f_a=160$~MHz corresponds to twice the intrinsic relaxation frequency since the phase-space is $\pi/2$-periodic in $\theta$. Eq.~(\ref{eq:Arrhenius}) reveals that an exponential linewidth broadening is expected as a function of $T^{-1}$ near a mode-transition, in contrast to single-mode predictions~\cite{Slavin2009}. Indeed, by estimating $\Delta E=52$~meV from a single experimental data at $303$~K~\cite{Muduli2012b,SuppMat}, we obtain the correct linewidth trend by simple evaluation of the Arrhenius equation [Fig.~\ref{fig3}(c)]. A better fit would involve possible changes in the fixed point position as a function of temperature~\cite{Banerjee2000} and temperature dependence of the coupling term~\cite{Heinonen2013}. However, these details are outside the scope of the present Letter.

In summary, we have presented an analytical description of the generation linewidth of a multi-mode STO, based on a multi-mode theory~\cite{Muduli2012,Muduli2012b,Heinonen2013}. The model equations demonstrate that mode-hopping events can dominate the generation linewidth and thus provide the main mechanism behind linewidth broadening. In particular, our results are in quantitative agreement with current and temperature dependent linewidths observed experimentally. Furthermore, we showed that the linewidth is dominated by high mode-hopping rates, providing means to determine the energy barrier separating both modes and the mechanism behind temperature dependent linewidth broadening. The presented results open up the possibility to study and determine intrinsic properties of multi-mode STOs.

Support from the Swedish Research Council (VR), the Swedish Foundation for Strategic Research (SSF), and the Knut and Alice Wallenberg Foundation is gratefully acknowledged. Argonne National Laboratory is operated under Contract No. DE-AC02-06CH11357 by UChicago Argonne, LLC.

\bibliographystyle{aipnum4-1}

\begin{thebibliography}{43}%
\makeatletter
\providecommand \@ifxundefined [1]{%
 \@ifx{#1\undefined}
}%
\providecommand \@ifnum [1]{%
 \ifnum #1\expandafter \@firstoftwo
 \else \expandafter \@secondoftwo
 \fi
}%
\providecommand \@ifx [1]{%
 \ifx #1\expandafter \@firstoftwo
 \else \expandafter \@secondoftwo
 \fi
}%
\providecommand \natexlab [1]{#1}%
\providecommand \enquote  [1]{``#1''}%
\providecommand \bibnamefont  [1]{#1}%
\providecommand \bibfnamefont [1]{#1}%
\providecommand \citenamefont [1]{#1}%
\providecommand \href@noop [0]{\@secondoftwo}%
\providecommand \href [0]{\begingroup \@sanitize@url \@href}%
\providecommand \@href[1]{\@@startlink{#1}\@@href}%
\providecommand \@@href[1]{\endgroup#1\@@endlink}%
\providecommand \@sanitize@url [0]{\catcode `\\12\catcode `\$12\catcode
  `\&12\catcode `\#12\catcode `\^12\catcode `\_12\catcode `\%12\relax}%
\providecommand \@@startlink[1]{}%
\providecommand \@@endlink[0]{}%
\providecommand \url  [0]{\begingroup\@sanitize@url \@url }%
\providecommand \@url [1]{\endgroup\@href {#1}{\urlprefix }}%
\providecommand \urlprefix  [0]{URL }%
\providecommand \Eprint [0]{\href }%
\providecommand \doibase [0]{http://dx.doi.org/}%
\providecommand \selectlanguage [0]{\@gobble}%
\providecommand \bibinfo  [0]{\@secondoftwo}%
\providecommand \bibfield  [0]{\@secondoftwo}%
\providecommand \translation [1]{[#1]}%
\providecommand \BibitemOpen [0]{}%
\providecommand \bibitemStop [0]{}%
\providecommand \bibitemNoStop [0]{.\EOS\space}%
\providecommand \EOS [0]{\spacefactor3000\relax}%
\providecommand \BibitemShut  [1]{\csname bibitem#1\endcsname}%
\let\auto@bib@innerbib\@empty
\bibitem [{\citenamefont {Slonczewski}(1996)}]{Slonczewski1996}%
  \BibitemOpen
  \bibfield  {author} {\bibinfo {author} {\bibfnamefont {J.~C.}\ \bibnamefont
  {Slonczewski}},\ }\href {\doibase DOI: 10.1016/0304-8853(96)00062-5}
  {\bibfield  {journal} {\bibinfo  {journal} {Journal of Magnetism and Magnetic
  Materials}\ }\textbf {\bibinfo {volume} {159}},\ \bibinfo {pages} {L1 }
  (\bibinfo {year} {1996})}\BibitemShut {NoStop}%
\bibitem [{\citenamefont {Berger}(1996)}]{Berger1996}%
  \BibitemOpen
  \bibfield  {author} {\bibinfo {author} {\bibfnamefont {L.}~\bibnamefont
  {Berger}},\ }\href {\doibase 10.1103/PhysRevB.54.9353} {\bibfield  {journal}
  {\bibinfo  {journal} {Phys. Rev. B}\ }\textbf {\bibinfo {volume} {54}},\
  \bibinfo {pages} {9353} (\bibinfo {year} {1996})}\BibitemShut {NoStop}%
\bibitem [{\citenamefont {Slonczewski}(1999)}]{Slonczewski1999}%
  \BibitemOpen
  \bibfield  {author} {\bibinfo {author} {\bibfnamefont {J.~C.}\ \bibnamefont
  {Slonczewski}},\ }\href {\doibase DOI: 10.1016/S0304-8853(99)00043-8}
  {\bibfield  {journal} {\bibinfo  {journal} {Journal of Magnetism and Magnetic
  Materials}\ }\textbf {\bibinfo {volume} {195}},\ \bibinfo {pages} {261 }
  (\bibinfo {year} {1999})}\BibitemShut {NoStop}%
\bibitem [{\citenamefont {Slavin}\ and\ \citenamefont
  {Tiberkevich}(2005)}]{Slavin2005}%
  \BibitemOpen
  \bibfield  {author} {\bibinfo {author} {\bibfnamefont {A.}~\bibnamefont
  {Slavin}}\ and\ \bibinfo {author} {\bibfnamefont {V.}~\bibnamefont
  {Tiberkevich}},\ }\href {\doibase 10.1103/PhysRevLett.95.237201} {\bibfield
  {journal} {\bibinfo  {journal} {Phys. Rev. Lett.}\ }\textbf {\bibinfo
  {volume} {95}},\ \bibinfo {pages} {237201} (\bibinfo {year}
  {2005})}\BibitemShut {NoStop}%
\bibitem [{\citenamefont {Bonetti}\ \emph {et~al.}(2010)\citenamefont
  {Bonetti}, \citenamefont {Tiberkevich}, \citenamefont {Consolo},
  \citenamefont {Finocchio}, \citenamefont {Muduli}, \citenamefont {Mancoff},
  \citenamefont {Slavin},\ and\ \citenamefont {\AA{}kerman}}]{Bonetti2010}%
  \BibitemOpen
  \bibfield  {author} {\bibinfo {author} {\bibfnamefont {S.}~\bibnamefont
  {Bonetti}}, \bibinfo {author} {\bibfnamefont {V.}~\bibnamefont
  {Tiberkevich}}, \bibinfo {author} {\bibfnamefont {G.}~\bibnamefont
  {Consolo}}, \bibinfo {author} {\bibfnamefont {G.}~\bibnamefont {Finocchio}},
  \bibinfo {author} {\bibfnamefont {P.}~\bibnamefont {Muduli}}, \bibinfo
  {author} {\bibfnamefont {F.}~\bibnamefont {Mancoff}}, \bibinfo {author}
  {\bibfnamefont {A.}~\bibnamefont {Slavin}}, \ and\ \bibinfo {author}
  {\bibfnamefont {J.}~\bibnamefont {\AA{}kerman}},\ }\href {\doibase
  10.1103/PhysRevLett.105.217204} {\bibfield  {journal} {\bibinfo  {journal}
  {Phys. Rev. Lett.}\ }\textbf {\bibinfo {volume} {105}},\ \bibinfo {pages}
  {217204} (\bibinfo {year} {2010})}\BibitemShut {NoStop}%
\bibitem [{\citenamefont {Madami}\ \emph {et~al.}(2011)\citenamefont {Madami},
  \citenamefont {Bonetti}, \citenamefont {Consolo}, \citenamefont {Tacchi},
  \citenamefont {Carlotti}, \citenamefont {Gubbiotti}, \citenamefont {Mancoff},
  \citenamefont {Yar},\ and\ \citenamefont {\AA{}kerman}}]{Madami2011}%
  \BibitemOpen
  \bibfield  {author} {\bibinfo {author} {\bibfnamefont {M.}~\bibnamefont
  {Madami}}, \bibinfo {author} {\bibfnamefont {S.}~\bibnamefont {Bonetti}},
  \bibinfo {author} {\bibfnamefont {G.}~\bibnamefont {Consolo}}, \bibinfo
  {author} {\bibfnamefont {S.}~\bibnamefont {Tacchi}}, \bibinfo {author}
  {\bibfnamefont {G.}~\bibnamefont {Carlotti}}, \bibinfo {author}
  {\bibfnamefont {G.}~\bibnamefont {Gubbiotti}}, \bibinfo {author}
  {\bibfnamefont {F.~B.}\ \bibnamefont {Mancoff}}, \bibinfo {author}
  {\bibfnamefont {M.~A.}\ \bibnamefont {Yar}}, \ and\ \bibinfo {author}
  {\bibfnamefont {J.}~\bibnamefont {\AA{}kerman}},\ }\href {\doibase
  10.138/nnano.2011.140} {\bibfield  {journal} {\bibinfo  {journal} {Nature
  Nanotechnology}\ }\textbf {\bibinfo {volume} {6}},\ \bibinfo {pages} {635}
  (\bibinfo {year} {2011})}\BibitemShut {NoStop}%
\bibitem [{\citenamefont {Bonetti}\ \emph {et~al.}(2012)\citenamefont
  {Bonetti}, \citenamefont {Puliafito}, \citenamefont {Consolo}, \citenamefont
  {Tiberkevich}, \citenamefont {Slavin},\ and\ \citenamefont
  {\AA{}kerman}}]{Bonetti2012}%
  \BibitemOpen
  \bibfield  {author} {\bibinfo {author} {\bibfnamefont {S.}~\bibnamefont
  {Bonetti}}, \bibinfo {author} {\bibfnamefont {V.}~\bibnamefont {Puliafito}},
  \bibinfo {author} {\bibfnamefont {G.}~\bibnamefont {Consolo}}, \bibinfo
  {author} {\bibfnamefont {V.~S.}\ \bibnamefont {Tiberkevich}}, \bibinfo
  {author} {\bibfnamefont {A.~N.}\ \bibnamefont {Slavin}}, \ and\ \bibinfo
  {author} {\bibfnamefont {J.}~\bibnamefont {\AA{}kerman}},\ }\href {\doibase
  10.1103/PhysRevB.85.174427} {\bibfield  {journal} {\bibinfo  {journal} {Phys.
  Rev. B}\ }\textbf {\bibinfo {volume} {85}},\ \bibinfo {pages} {174427}
  (\bibinfo {year} {2012})}\BibitemShut {NoStop}%
\bibitem [{\citenamefont {Dumas}\ \emph {et~al.}(2013)\citenamefont {Dumas},
  \citenamefont {Iacocca}, \citenamefont {Bonetti}, \citenamefont {Sani},
  \citenamefont {Mohseni}, \citenamefont {Eklund}, \citenamefont {Persson},
  \citenamefont {Heinonen},\ and\ \citenamefont {\AA{}kerman}}]{Dumas2013}%
  \BibitemOpen
  \bibfield  {author} {\bibinfo {author} {\bibfnamefont {R.~K.}\ \bibnamefont
  {Dumas}}, \bibinfo {author} {\bibfnamefont {E.}~\bibnamefont {Iacocca}},
  \bibinfo {author} {\bibfnamefont {S.}~\bibnamefont {Bonetti}}, \bibinfo
  {author} {\bibfnamefont {S.~R.}\ \bibnamefont {Sani}}, \bibinfo {author}
  {\bibfnamefont {S.~M.}\ \bibnamefont {Mohseni}}, \bibinfo {author}
  {\bibfnamefont {A.}~\bibnamefont {Eklund}}, \bibinfo {author} {\bibfnamefont
  {J.}~\bibnamefont {Persson}}, \bibinfo {author} {\bibfnamefont
  {O.}~\bibnamefont {Heinonen}}, \ and\ \bibinfo {author} {\bibfnamefont
  {J.}~\bibnamefont {\AA{}kerman}},\ }\href {\doibase
  10.1103/PhysRevLett.110.257202} {\bibfield  {journal} {\bibinfo  {journal}
  {Phys. Rev. Lett.}\ }\textbf {\bibinfo {volume} {110}},\ \bibinfo {pages}
  {257202} (\bibinfo {year} {2013})}\BibitemShut {NoStop}%
\bibitem [{\citenamefont {Pribiag}\ \emph {et~al.}(2007)\citenamefont
  {Pribiag}, \citenamefont {Krivorotov}, \citenamefont {Fuchs}, \citenamefont
  {Braganca}, \citenamefont {Ozatay}, \citenamefont {Sankey}, \citenamefont
  {Ralph},\ and\ \citenamefont {Buhrman}}]{Pribiag2007}%
  \BibitemOpen
  \bibfield  {author} {\bibinfo {author} {\bibfnamefont {V.}~\bibnamefont
  {Pribiag}}, \bibinfo {author} {\bibfnamefont {I.~N.}\ \bibnamefont
  {Krivorotov}}, \bibinfo {author} {\bibfnamefont {G.~D.}\ \bibnamefont
  {Fuchs}}, \bibinfo {author} {\bibfnamefont {P.~M.}\ \bibnamefont {Braganca}},
  \bibinfo {author} {\bibfnamefont {O.}~\bibnamefont {Ozatay}}, \bibinfo
  {author} {\bibfnamefont {J.~C.}\ \bibnamefont {Sankey}}, \bibinfo {author}
  {\bibfnamefont {D.~C.}\ \bibnamefont {Ralph}}, \ and\ \bibinfo {author}
  {\bibfnamefont {R.~A.}\ \bibnamefont {Buhrman}},\ }\href@noop {} {\bibfield
  {journal} {\bibinfo  {journal} {Nature Physics}\ }\textbf {\bibinfo {volume}
  {3}},\ \bibinfo {pages} {489 } (\bibinfo {year} {2007})}\BibitemShut
  {NoStop}%
\bibitem [{\citenamefont {Devolder}\ \emph {et~al.}(2010)\citenamefont
  {Devolder}, \citenamefont {Kim}, \citenamefont {Manfrini}, \citenamefont {van
  Roy}, \citenamefont {Lagae},\ and\ \citenamefont {Chappert}}]{Devolder2010}%
  \BibitemOpen
  \bibfield  {author} {\bibinfo {author} {\bibfnamefont {T.}~\bibnamefont
  {Devolder}}, \bibinfo {author} {\bibfnamefont {J.-V.}\ \bibnamefont {Kim}},
  \bibinfo {author} {\bibfnamefont {M.}~\bibnamefont {Manfrini}}, \bibinfo
  {author} {\bibfnamefont {W.}~\bibnamefont {van Roy}}, \bibinfo {author}
  {\bibfnamefont {L.}~\bibnamefont {Lagae}}, \ and\ \bibinfo {author}
  {\bibfnamefont {C.}~\bibnamefont {Chappert}},\ }\href {\doibase
  10.1063/1.3478843} {\bibfield  {journal} {\bibinfo  {journal} {Applied
  Physics Letters}\ }\textbf {\bibinfo {volume} {97}},\ \bibinfo {eid} {072512}
  (\bibinfo {year} {2010})}\BibitemShut {NoStop}%
\bibitem [{\citenamefont {Dussaux}\ \emph {et~al.}(2010)\citenamefont
  {Dussaux}, \citenamefont {Georges}, \citenamefont {Grollier}, \citenamefont
  {Cros}, \citenamefont {Khvalkovskiy}, \citenamefont {Fukushima},
  \citenamefont {Konoto}, \citenamefont {Kubota}, \citenamefont {Yakushiji},
  \citenamefont {Yuasa}, \citenamefont {Zvezdin}, \citenamefont {Ando},\ and\
  \citenamefont {Fert}}]{Dussaux2010}%
  \BibitemOpen
  \bibfield  {author} {\bibinfo {author} {\bibfnamefont {A.}~\bibnamefont
  {Dussaux}}, \bibinfo {author} {\bibfnamefont {B.}~\bibnamefont {Georges}},
  \bibinfo {author} {\bibfnamefont {J.}~\bibnamefont {Grollier}}, \bibinfo
  {author} {\bibfnamefont {V.}~\bibnamefont {Cros}}, \bibinfo {author}
  {\bibfnamefont {A.~V.}\ \bibnamefont {Khvalkovskiy}}, \bibinfo {author}
  {\bibfnamefont {A.}~\bibnamefont {Fukushima}}, \bibinfo {author}
  {\bibfnamefont {M.}~\bibnamefont {Konoto}}, \bibinfo {author} {\bibfnamefont
  {H.}~\bibnamefont {Kubota}}, \bibinfo {author} {\bibfnamefont
  {K.}~\bibnamefont {Yakushiji}}, \bibinfo {author} {\bibfnamefont
  {S.}~\bibnamefont {Yuasa}}, \bibinfo {author} {\bibfnamefont {A.~K.}\
  \bibnamefont {Zvezdin}}, \bibinfo {author} {\bibfnamefont {K.}~\bibnamefont
  {Ando}}, \ and\ \bibinfo {author} {\bibfnamefont {A.}~\bibnamefont {Fert}},\
  }\href@noop {} {\bibfield  {journal} {\bibinfo  {journal} {Nature
  Communications}\ }\textbf {\bibinfo {volume} {1}},\ \bibinfo {pages} {8}
  (\bibinfo {year} {2010})}\BibitemShut {NoStop}%
\bibitem [{\citenamefont {Petit-Watelot}\ \emph {et~al.}(2012)\citenamefont
  {Petit-Watelot}, \citenamefont {Kim}, \citenamefont {Otxoa}, \citenamefont
  {Bouzehouane}, \citenamefont {Grollier}, \citenamefont {Vansteenkiste},
  \citenamefont {Van~de Wiele}, \citenamefont {Cros},\ and\ \citenamefont
  {Devolder}}]{Petit2012}%
  \BibitemOpen
  \bibfield  {author} {\bibinfo {author} {\bibfnamefont {S.}~\bibnamefont
  {Petit-Watelot}}, \bibinfo {author} {\bibfnamefont {A.}~\bibnamefont {Kim},
  \bibfnamefont {Joo-Von~Ruotolo}}, \bibinfo {author} {\bibfnamefont {R.~M.}\
  \bibnamefont {Otxoa}}, \bibinfo {author} {\bibfnamefont {K.}~\bibnamefont
  {Bouzehouane}}, \bibinfo {author} {\bibfnamefont {J.}~\bibnamefont
  {Grollier}}, \bibinfo {author} {\bibfnamefont {A.}~\bibnamefont
  {Vansteenkiste}}, \bibinfo {author} {\bibfnamefont {B.}~\bibnamefont {Van~de
  Wiele}}, \bibinfo {author} {\bibfnamefont {V.}~\bibnamefont {Cros}}, \ and\
  \bibinfo {author} {\bibfnamefont {T.}~\bibnamefont {Devolder}},\ }\href
  {\doibase 10.1038/nphys2362} {\bibfield  {journal} {\bibinfo  {journal}
  {Nature Physics}\ }\textbf {\bibinfo {volume} {8}},\ \bibinfo {pages} {682 }
  (\bibinfo {year} {2012})}\BibitemShut {NoStop}%
\bibitem [{\citenamefont {Hoefer}, \citenamefont {Silva},\ and\ \citenamefont
  {Keller}(2010)}]{Hoefer2010}%
  \BibitemOpen
  \bibfield  {author} {\bibinfo {author} {\bibfnamefont {M.~A.}\ \bibnamefont
  {Hoefer}}, \bibinfo {author} {\bibfnamefont {T.~J.}\ \bibnamefont {Silva}}, \
  and\ \bibinfo {author} {\bibfnamefont {M.~W.}\ \bibnamefont {Keller}},\
  }\href {\doibase 10.1103/PhysRevB.82.054432} {\bibfield  {journal} {\bibinfo
  {journal} {Phys. Rev. B}\ }\textbf {\bibinfo {volume} {82}},\ \bibinfo
  {pages} {054432} (\bibinfo {year} {2010})}\BibitemShut {NoStop}%
\bibitem [{\citenamefont {Hoefer}, \citenamefont {Sommacal},\ and\
  \citenamefont {Silva}(2012)}]{Hoefer2012}%
  \BibitemOpen
  \bibfield  {author} {\bibinfo {author} {\bibfnamefont {M.~A.}\ \bibnamefont
  {Hoefer}}, \bibinfo {author} {\bibfnamefont {M.}~\bibnamefont {Sommacal}}, \
  and\ \bibinfo {author} {\bibfnamefont {T.~J.}\ \bibnamefont {Silva}},\ }\href
  {\doibase 10.1103/PhysRevB.85.214433} {\bibfield  {journal} {\bibinfo
  {journal} {Phys. Rev. B}\ }\textbf {\bibinfo {volume} {85}},\ \bibinfo
  {pages} {214433} (\bibinfo {year} {2012})}\BibitemShut {NoStop}%
\bibitem [{\citenamefont {Mohseni}\ \emph
  {et~al.}(2013{\natexlab{a}})\citenamefont {Mohseni}, \citenamefont {Sani},
  \citenamefont {Persson}, \citenamefont {Nguyen}, \citenamefont {Chung},
  \citenamefont {Pogoryelov}, \citenamefont {Muduli}, \citenamefont {Iacocca},
  \citenamefont {Eklund}, \citenamefont {Dumas}, \citenamefont {Bonetti},
  \citenamefont {Deac}, \citenamefont {Hoefer},\ and\ \citenamefont
  {\AA{}kerman}}]{Mohseni2013}%
  \BibitemOpen
  \bibfield  {author} {\bibinfo {author} {\bibfnamefont {S.~M.}\ \bibnamefont
  {Mohseni}}, \bibinfo {author} {\bibfnamefont {S.~R.}\ \bibnamefont {Sani}},
  \bibinfo {author} {\bibfnamefont {J.}~\bibnamefont {Persson}}, \bibinfo
  {author} {\bibfnamefont {T.~N.~A.}\ \bibnamefont {Nguyen}}, \bibinfo {author}
  {\bibfnamefont {S.}~\bibnamefont {Chung}}, \bibinfo {author} {\bibfnamefont
  {Y.}~\bibnamefont {Pogoryelov}}, \bibinfo {author} {\bibfnamefont {P.~K.}\
  \bibnamefont {Muduli}}, \bibinfo {author} {\bibfnamefont {E.}~\bibnamefont
  {Iacocca}}, \bibinfo {author} {\bibfnamefont {A.}~\bibnamefont {Eklund}},
  \bibinfo {author} {\bibfnamefont {R.~K.}\ \bibnamefont {Dumas}}, \bibinfo
  {author} {\bibfnamefont {S.}~\bibnamefont {Bonetti}}, \bibinfo {author}
  {\bibfnamefont {A.}~\bibnamefont {Deac}}, \bibinfo {author} {\bibfnamefont
  {M.~A.}\ \bibnamefont {Hoefer}}, \ and\ \bibinfo {author} {\bibfnamefont
  {J.}~\bibnamefont {\AA{}kerman}},\ }\href {\doibase 10.1126/science.1230155}
  {\bibfield  {journal} {\bibinfo  {journal} {Science}\ }\textbf {\bibinfo
  {volume} {339}},\ \bibinfo {pages} {1295} (\bibinfo {year}
  {2013}{\natexlab{a}})}\BibitemShut {NoStop}%
\bibitem [{\citenamefont {Mohseni}\ \emph
  {et~al.}(2013{\natexlab{b}})\citenamefont {Mohseni}, \citenamefont {Sani},
  \citenamefont {Dumas}, \citenamefont {Persson}, \citenamefont {Nguyen},
  \citenamefont {Chung}, \citenamefont {Pogoryelov}, \citenamefont {Muduli},
  \citenamefont {Iacocca}, \citenamefont {Eklund},\ and\ \citenamefont
  {\AA{}kerman}}]{Mohseni2013b}%
  \BibitemOpen
  \bibfield  {author} {\bibinfo {author} {\bibfnamefont {S.}~\bibnamefont
  {Mohseni}}, \bibinfo {author} {\bibfnamefont {S.}~\bibnamefont {Sani}},
  \bibinfo {author} {\bibfnamefont {R.}~\bibnamefont {Dumas}}, \bibinfo
  {author} {\bibfnamefont {J.}~\bibnamefont {Persson}}, \bibinfo {author}
  {\bibfnamefont {T.~A.}\ \bibnamefont {Nguyen}}, \bibinfo {author}
  {\bibfnamefont {S.}~\bibnamefont {Chung}}, \bibinfo {author} {\bibfnamefont
  {Y.}~\bibnamefont {Pogoryelov}}, \bibinfo {author} {\bibfnamefont
  {P.}~\bibnamefont {Muduli}}, \bibinfo {author} {\bibfnamefont
  {E.}~\bibnamefont {Iacocca}}, \bibinfo {author} {\bibfnamefont
  {A.}~\bibnamefont {Eklund}}, \ and\ \bibinfo {author} {\bibfnamefont
  {J.}~\bibnamefont {\AA{}kerman}},\ }\href {\doibase DOI: 10.1016/j.physb.2013.10.023} {\bibfield  {journal}
  {\bibinfo  {journal} {Physica B: Condensed Matter}\ ,\ } (\bibinfo {year}
  {2013}{\natexlab{b}})}\BibitemShut {NoStop}%
\bibitem [{\citenamefont {Silva}\ and\ \citenamefont
  {Rippard}(2008)}]{Silva2008}%
  \BibitemOpen
  \bibfield  {author} {\bibinfo {author} {\bibfnamefont {T.}~\bibnamefont
  {Silva}}\ and\ \bibinfo {author} {\bibfnamefont {W.}~\bibnamefont
  {Rippard}},\ }\href {\doibase DOI: 10.1016/j.jmmm.2007.12.022} {\bibfield
  {journal} {\bibinfo  {journal} {Journal of Magnetism and Magnetic Materials}\
  }\textbf {\bibinfo {volume} {320}},\ \bibinfo {pages} {1260 } (\bibinfo
  {year} {2008})}\BibitemShut {NoStop}%
\bibitem [{\citenamefont {Ralph}\ and\ \citenamefont
  {Stiles}(2008)}]{Ralph2008}%
  \BibitemOpen
  \bibfield  {author} {\bibinfo {author} {\bibfnamefont {D.}~\bibnamefont
  {Ralph}}\ and\ \bibinfo {author} {\bibfnamefont {M.}~\bibnamefont {Stiles}},\
  }\href {\doibase DOI: 10.1016/j.jmmm.2007.12.019} {\bibfield  {journal}
  {\bibinfo  {journal} {Journal of Magnetism and Magnetic Materials}\ }\textbf
  {\bibinfo {volume} {320}},\ \bibinfo {pages} {1190 } (\bibinfo {year}
  {2008})}\BibitemShut {NoStop}%
\bibitem [{\citenamefont {Sani}\ \emph {et~al.}(2013)\citenamefont {Sani},
  \citenamefont {D\"{u}rrenfeld}, \citenamefont {Mohseni}, \citenamefont
  {Chung},\ and\ \citenamefont {\AA{}kerman}}]{Sani2013}%
  \BibitemOpen
  \bibfield  {author} {\bibinfo {author} {\bibfnamefont {S.}~\bibnamefont
  {Sani}}, \bibinfo {author} {\bibfnamefont {P.}~\bibnamefont
  {D\"{u}rrenfeld}}, \bibinfo {author} {\bibfnamefont {S.}~\bibnamefont
  {Mohseni}}, \bibinfo {author} {\bibfnamefont {S.}~\bibnamefont {Chung}}, \
  and\ \bibinfo {author} {\bibfnamefont {J.}~\bibnamefont {\AA{}kerman}},\
  }\href {\doibase 10.1109/TMAG.2013.2250931} {\bibfield  {journal} {\bibinfo
  {journal} {Magnetics, IEEE Transactions on}\ }\textbf {\bibinfo {volume}
  {49}},\ \bibinfo {pages} {4331} (\bibinfo {year} {2013})}\BibitemShut
  {NoStop}%
\bibitem [{\citenamefont {Baibich}\ \emph {et~al.}(1988)\citenamefont
  {Baibich}, \citenamefont {Broto}, \citenamefont {Fert}, \citenamefont
  {Van~Dau}, \citenamefont {Petroff}, \citenamefont {Etienne}, \citenamefont
  {Creuzet}, \citenamefont {Friederich},\ and\ \citenamefont
  {Chazelas}}]{Baibich1988}%
  \BibitemOpen
  \bibfield  {author} {\bibinfo {author} {\bibfnamefont {M.~N.}\ \bibnamefont
  {Baibich}}, \bibinfo {author} {\bibfnamefont {J.~M.}\ \bibnamefont {Broto}},
  \bibinfo {author} {\bibfnamefont {A.}~\bibnamefont {Fert}}, \bibinfo {author}
  {\bibfnamefont {F.~N.}\ \bibnamefont {Van~Dau}}, \bibinfo {author}
  {\bibfnamefont {F.}~\bibnamefont {Petroff}}, \bibinfo {author} {\bibfnamefont
  {P.}~\bibnamefont {Etienne}}, \bibinfo {author} {\bibfnamefont
  {G.}~\bibnamefont {Creuzet}}, \bibinfo {author} {\bibfnamefont
  {A.}~\bibnamefont {Friederich}}, \ and\ \bibinfo {author} {\bibfnamefont
  {J.}~\bibnamefont {Chazelas}},\ }\href {\doibase 10.1103/PhysRevLett.61.2472}
  {\bibfield  {journal} {\bibinfo  {journal} {Phys. Rev. Lett.}\ }\textbf
  {\bibinfo {volume} {61}},\ \bibinfo {pages} {2472} (\bibinfo {year}
  {1988})}\BibitemShut {NoStop}%
\bibitem [{\citenamefont {Binasch}\ \emph {et~al.}(1989)\citenamefont
  {Binasch}, \citenamefont {Gr\"unberg}, \citenamefont {Saurenbach},\ and\
  \citenamefont {Zinn}}]{Binasch1989}%
  \BibitemOpen
  \bibfield  {author} {\bibinfo {author} {\bibfnamefont {G.}~\bibnamefont
  {Binasch}}, \bibinfo {author} {\bibfnamefont {P.}~\bibnamefont {Gr\"unberg}},
  \bibinfo {author} {\bibfnamefont {F.}~\bibnamefont {Saurenbach}}, \ and\
  \bibinfo {author} {\bibfnamefont {W.}~\bibnamefont {Zinn}},\ }\href {\doibase
  10.1103/PhysRevB.39.4828} {\bibfield  {journal} {\bibinfo  {journal} {Phys.
  Rev. B}\ }\textbf {\bibinfo {volume} {39}},\ \bibinfo {pages} {4828}
  (\bibinfo {year} {1989})}\BibitemShut {NoStop}%
\bibitem [{\citenamefont {Tsoi}\ \emph {et~al.}(1998)\citenamefont {Tsoi},
  \citenamefont {Jansen}, \citenamefont {Bass}, \citenamefont {Chiang},
  \citenamefont {Seck}, \citenamefont {Tsoi},\ and\ \citenamefont
  {Wyder}}]{Tsoi1998}%
  \BibitemOpen
  \bibfield  {author} {\bibinfo {author} {\bibfnamefont {M.}~\bibnamefont
  {Tsoi}}, \bibinfo {author} {\bibfnamefont {A.~G.~M.}\ \bibnamefont {Jansen}},
  \bibinfo {author} {\bibfnamefont {J.}~\bibnamefont {Bass}}, \bibinfo {author}
  {\bibfnamefont {W.-C.}\ \bibnamefont {Chiang}}, \bibinfo {author}
  {\bibfnamefont {M.}~\bibnamefont {Seck}}, \bibinfo {author} {\bibfnamefont
  {V.}~\bibnamefont {Tsoi}}, \ and\ \bibinfo {author} {\bibfnamefont
  {P.}~\bibnamefont {Wyder}},\ }\href {\doibase 10.1103/PhysRevLett.80.4281}
  {\bibfield  {journal} {\bibinfo  {journal} {Phys. Rev. Lett.}\ }\textbf
  {\bibinfo {volume} {80}},\ \bibinfo {pages} {4281} (\bibinfo {year}
  {1998})}\BibitemShut {NoStop}%
\bibitem [{\citenamefont {Julliere}(1975)}]{Julliere1975}%
  \BibitemOpen
  \bibfield  {author} {\bibinfo {author} {\bibfnamefont {M.}~\bibnamefont
  {Julliere}},\ }\href
  {http://linkinghub.elsevier.com/retrieve/pii/0375960175901747} {\bibfield
  {journal} {\bibinfo  {journal} {Physics Letters A}\ }\textbf {\bibinfo
  {volume} {54}},\ \bibinfo {pages} {225} (\bibinfo {year} {1975})}\BibitemShut
  {NoStop}%
\bibitem [{\citenamefont {Yuasa}\ \emph {et~al.}(2004)\citenamefont {Yuasa},
  \citenamefont {Nagahama}, \citenamefont {Fukushima}, \citenamefont {Suzuki},\
  and\ \citenamefont {Ando}}]{Yuasa2004}%
  \BibitemOpen
  \bibfield  {author} {\bibinfo {author} {\bibfnamefont {S.}~\bibnamefont
  {Yuasa}}, \bibinfo {author} {\bibfnamefont {T.}~\bibnamefont {Nagahama}},
  \bibinfo {author} {\bibfnamefont {A.}~\bibnamefont {Fukushima}}, \bibinfo
  {author} {\bibfnamefont {Y.}~\bibnamefont {Suzuki}}, \ and\ \bibinfo {author}
  {\bibfnamefont {K.}~\bibnamefont {Ando}},\ }\href@noop {} {\bibfield
  {journal} {\bibinfo  {journal} {Nature Materials}\ }\textbf {\bibinfo
  {volume} {3}},\ \bibinfo {pages} {868 } (\bibinfo {year} {2004})}\BibitemShut
  {NoStop}%
\bibitem [{\citenamefont {Houssameddine}\ \emph {et~al.}(2008)\citenamefont
  {Houssameddine}, \citenamefont {Florez}, \citenamefont {Katine},
  \citenamefont {Michel}, \citenamefont {Ebels}, \citenamefont {Mauri},
  \citenamefont {Ozatay}, \citenamefont {Delaet}, \citenamefont {Viala},
  \citenamefont {Folks}, \citenamefont {Terris},\ and\ \citenamefont
  {Cyrille}}]{Houssameddine2008}%
  \BibitemOpen
  \bibfield  {author} {\bibinfo {author} {\bibfnamefont {D.}~\bibnamefont
  {Houssameddine}}, \bibinfo {author} {\bibfnamefont {S.~H.}\ \bibnamefont
  {Florez}}, \bibinfo {author} {\bibfnamefont {J.~A.}\ \bibnamefont {Katine}},
  \bibinfo {author} {\bibfnamefont {J.-P.}\ \bibnamefont {Michel}}, \bibinfo
  {author} {\bibfnamefont {U.}~\bibnamefont {Ebels}}, \bibinfo {author}
  {\bibfnamefont {D.}~\bibnamefont {Mauri}}, \bibinfo {author} {\bibfnamefont
  {O.}~\bibnamefont {Ozatay}}, \bibinfo {author} {\bibfnamefont
  {B.}~\bibnamefont {Delaet}}, \bibinfo {author} {\bibfnamefont
  {B.}~\bibnamefont {Viala}}, \bibinfo {author} {\bibfnamefont
  {L.}~\bibnamefont {Folks}}, \bibinfo {author} {\bibfnamefont {B.~D.}\
  \bibnamefont {Terris}}, \ and\ \bibinfo {author} {\bibfnamefont {M.-C.}\
  \bibnamefont {Cyrille}},\ }\href {\doibase 10.1063/1.2956418} {\bibfield
  {journal} {\bibinfo  {journal} {Applied Physics Letters}\ }\textbf {\bibinfo
  {volume} {93}},\ \bibinfo {eid} {022505} (\bibinfo {year}
  {2008})}\BibitemShut {NoStop}%
\bibitem [{\citenamefont {Rezende}, \citenamefont {de~Aguiar},\ and\
  \citenamefont {Azevedo}(2005)}]{Rezende2005}%
  \BibitemOpen
  \bibfield  {author} {\bibinfo {author} {\bibfnamefont {S.~M.}\ \bibnamefont
  {Rezende}}, \bibinfo {author} {\bibfnamefont {F.~M.}\ \bibnamefont
  {de~Aguiar}}, \ and\ \bibinfo {author} {\bibfnamefont {A.}~\bibnamefont
  {Azevedo}},\ }\href {\doibase 10.1103/PhysRevLett.94.037202} {\bibfield
  {journal} {\bibinfo  {journal} {Phys. Rev. Lett.}\ }\textbf {\bibinfo
  {volume} {94}},\ \bibinfo {pages} {037202} (\bibinfo {year}
  {2005})}\BibitemShut {NoStop}%
\bibitem [{\citenamefont {Slavin}\ and\ \citenamefont
  {Tiberkevich}(2009)}]{Slavin2009}%
  \BibitemOpen
  \bibfield  {author} {\bibinfo {author} {\bibfnamefont {A.}~\bibnamefont
  {Slavin}}\ and\ \bibinfo {author} {\bibfnamefont {V.}~\bibnamefont
  {Tiberkevich}},\ }\href {\doibase 10.1109/TMAG.2008.2009935} {\bibfield
  {journal} {\bibinfo  {journal} {Magnetics, IEEE Transactions on}\ }\textbf
  {\bibinfo {volume} {45}},\ \bibinfo {pages} {1875 } (\bibinfo {year}
  {2009})}\BibitemShut {NoStop}%
\bibitem [{\citenamefont {Kiselev}\ \emph {et~al.}(2004)\citenamefont
  {Kiselev}, \citenamefont {Sankey}, \citenamefont {Krivorotov}, \citenamefont
  {Emley}, \citenamefont {Rinkoski}, \citenamefont {Perez}, \citenamefont
  {Buhrman},\ and\ \citenamefont {Ralph}}]{Kiselev2004}%
  \BibitemOpen
  \bibfield  {author} {\bibinfo {author} {\bibfnamefont {S.~I.}\ \bibnamefont
  {Kiselev}}, \bibinfo {author} {\bibfnamefont {J.~C.}\ \bibnamefont {Sankey}},
  \bibinfo {author} {\bibfnamefont {I.~N.}\ \bibnamefont {Krivorotov}},
  \bibinfo {author} {\bibfnamefont {N.~C.}\ \bibnamefont {Emley}}, \bibinfo
  {author} {\bibfnamefont {M.}~\bibnamefont {Rinkoski}}, \bibinfo {author}
  {\bibfnamefont {C.}~\bibnamefont {Perez}}, \bibinfo {author} {\bibfnamefont
  {R.~A.}\ \bibnamefont {Buhrman}}, \ and\ \bibinfo {author} {\bibfnamefont
  {D.~C.}\ \bibnamefont {Ralph}},\ }\href {\doibase
  10.1103/PhysRevLett.93.036601} {\bibfield  {journal} {\bibinfo  {journal}
  {Phys. Rev. Lett.}\ }\textbf {\bibinfo {volume} {93}},\ \bibinfo {pages}
  {036601} (\bibinfo {year} {2004})}\BibitemShut {NoStop}%
\bibitem [{\citenamefont {Sankey}\ \emph {et~al.}(2005)\citenamefont {Sankey},
  \citenamefont {Krivorotov}, \citenamefont {Kiselev}, \citenamefont
  {Braganca}, \citenamefont {Emley}, \citenamefont {Buhrman},\ and\
  \citenamefont {Ralph}}]{Sankey2005}%
  \BibitemOpen
  \bibfield  {author} {\bibinfo {author} {\bibfnamefont {J.~C.}\ \bibnamefont
  {Sankey}}, \bibinfo {author} {\bibfnamefont {I.~N.}\ \bibnamefont
  {Krivorotov}}, \bibinfo {author} {\bibfnamefont {S.~I.}\ \bibnamefont
  {Kiselev}}, \bibinfo {author} {\bibfnamefont {P.~M.}\ \bibnamefont
  {Braganca}}, \bibinfo {author} {\bibfnamefont {N.~C.}\ \bibnamefont {Emley}},
  \bibinfo {author} {\bibfnamefont {R.~A.}\ \bibnamefont {Buhrman}}, \ and\
  \bibinfo {author} {\bibfnamefont {D.~C.}\ \bibnamefont {Ralph}},\ }\href
  {\doibase 10.1103/PhysRevB.72.224427} {\bibfield  {journal} {\bibinfo
  {journal} {Phys. Rev. B}\ }\textbf {\bibinfo {volume} {72}},\ \bibinfo
  {pages} {224427} (\bibinfo {year} {2005})}\BibitemShut {NoStop}%
\bibitem [{\citenamefont {Zeng}\ \emph {et~al.}(2010)\citenamefont {Zeng},
  \citenamefont {Cheung}, \citenamefont {Jiang}, \citenamefont {Krivorotov},
  \citenamefont {Katine}, \citenamefont {Tiberkevich},\ and\ \citenamefont
  {Slavin}}]{Zeng2010}%
  \BibitemOpen
  \bibfield  {author} {\bibinfo {author} {\bibfnamefont {Z.}~\bibnamefont
  {Zeng}}, \bibinfo {author} {\bibfnamefont {K.~H.}\ \bibnamefont {Cheung}},
  \bibinfo {author} {\bibfnamefont {H.~W.}\ \bibnamefont {Jiang}}, \bibinfo
  {author} {\bibfnamefont {I.~N.}\ \bibnamefont {Krivorotov}}, \bibinfo
  {author} {\bibfnamefont {J.~A.}\ \bibnamefont {Katine}}, \bibinfo {author}
  {\bibfnamefont {V.}~\bibnamefont {Tiberkevich}}, \ and\ \bibinfo {author}
  {\bibfnamefont {A.}~\bibnamefont {Slavin}},\ }\href {\doibase
  10.1103/PhysRevB.82.100410} {\bibfield  {journal} {\bibinfo  {journal} {Phys.
  Rev. B}\ }\textbf {\bibinfo {volume} {82}},\ \bibinfo {pages} {100410}
  (\bibinfo {year} {2010})}\BibitemShut {NoStop}%
\bibitem [{\citenamefont {Krivorotov}\ \emph {et~al.}(2008)\citenamefont
  {Krivorotov}, \citenamefont {Emley}, \citenamefont {Buhrman},\ and\
  \citenamefont {Ralph}}]{Krivorotov2008}%
  \BibitemOpen
  \bibfield  {author} {\bibinfo {author} {\bibfnamefont {I.~N.}\ \bibnamefont
  {Krivorotov}}, \bibinfo {author} {\bibfnamefont {N.~C.}\ \bibnamefont
  {Emley}}, \bibinfo {author} {\bibfnamefont {R.~A.}\ \bibnamefont {Buhrman}},
  \ and\ \bibinfo {author} {\bibfnamefont {D.~C.}\ \bibnamefont {Ralph}},\
  }\href {\doibase 10.1103/PhysRevB.77.054440} {\bibfield  {journal} {\bibinfo
  {journal} {Phys. Rev. B}\ }\textbf {\bibinfo {volume} {77}},\ \bibinfo
  {pages} {054440} (\bibinfo {year} {2008})}\BibitemShut {NoStop}%
\bibitem [{\citenamefont {Muduli}, \citenamefont {Heinonen},\ and\
  \citenamefont {\AA{}kerman}(2012{\natexlab{a}})}]{Muduli2012}%
  \BibitemOpen
  \bibfield  {author} {\bibinfo {author} {\bibfnamefont {P.~K.}\ \bibnamefont
  {Muduli}}, \bibinfo {author} {\bibfnamefont {O.~G.}\ \bibnamefont
  {Heinonen}}, \ and\ \bibinfo {author} {\bibfnamefont {J.}~\bibnamefont
  {\AA{}kerman}},\ }\href {\doibase 10.1103/PhysRevLett.108.207203} {\bibfield
  {journal} {\bibinfo  {journal} {Phys. Rev. Lett.}\ }\textbf {\bibinfo
  {volume} {108}},\ \bibinfo {pages} {207203} (\bibinfo {year}
  {2012}{\natexlab{a}})}\BibitemShut {NoStop}%
\bibitem [{\citenamefont {Muduli}, \citenamefont {Heinonen},\ and\
  \citenamefont {\AA{}kerman}(2012{\natexlab{b}})}]{Muduli2012b}%
  \BibitemOpen
  \bibfield  {author} {\bibinfo {author} {\bibfnamefont {P.~K.}\ \bibnamefont
  {Muduli}}, \bibinfo {author} {\bibfnamefont {O.~G.}\ \bibnamefont
  {Heinonen}}, \ and\ \bibinfo {author} {\bibfnamefont {J.}~\bibnamefont
  {\AA{}kerman}},\ }\href {\doibase 10.1103/PhysRevB.86.174408} {\bibfield
  {journal} {\bibinfo  {journal} {Phys. Rev. B}\ }\textbf {\bibinfo {volume}
  {86}},\ \bibinfo {pages} {174408} (\bibinfo {year}
  {2012}{\natexlab{b}})}\BibitemShut {NoStop}%
\bibitem [{\citenamefont {Heinonen}\ \emph {et~al.}(2013)\citenamefont
  {Heinonen}, \citenamefont {Muduli}, \citenamefont {Iacocca},\ and\
  \citenamefont {\AA{}kerman}}]{Heinonen2013}%
  \BibitemOpen
  \bibfield  {author} {\bibinfo {author} {\bibfnamefont {O.}~\bibnamefont
  {Heinonen}}, \bibinfo {author} {\bibfnamefont {P.}~\bibnamefont {Muduli}},
  \bibinfo {author} {\bibfnamefont {E.}~\bibnamefont {Iacocca}}, \ and\
  \bibinfo {author} {\bibfnamefont {J.}~\bibnamefont {\AA{}kerman}},\ }\href
  {\doibase 10.1109/TMAG.2013.2242866} {\bibfield  {journal} {\bibinfo
  {journal} {Magnetics, IEEE Transactions on}\ }\textbf {\bibinfo {volume}
  {49}},\ \bibinfo {pages} {4398} (\bibinfo {year} {2013})}\BibitemShut
  {NoStop}%
\bibitem [{\citenamefont {Beri}\ \emph {et~al.}(2008)\citenamefont {Beri},
  \citenamefont {Gelens}, \citenamefont {Mestre}, \citenamefont {Van~der
  Sande}, \citenamefont {Verschaffelt}, \citenamefont {Scir\`e}, \citenamefont
  {Mezosi}, \citenamefont {Sorel},\ and\ \citenamefont {Danckaert}}]{Beri2008}%
  \BibitemOpen
  \bibfield  {author} {\bibinfo {author} {\bibfnamefont {S.}~\bibnamefont
  {Beri}}, \bibinfo {author} {\bibfnamefont {L.}~\bibnamefont {Gelens}},
  \bibinfo {author} {\bibfnamefont {M.}~\bibnamefont {Mestre}}, \bibinfo
  {author} {\bibfnamefont {G.}~\bibnamefont {Van~der Sande}}, \bibinfo {author}
  {\bibfnamefont {G.}~\bibnamefont {Verschaffelt}}, \bibinfo {author}
  {\bibfnamefont {A.}~\bibnamefont {Scir\'e}}, \bibinfo {author} {\bibfnamefont
  {G.}~\bibnamefont {Mezosi}}, \bibinfo {author} {\bibfnamefont
  {M.}~\bibnamefont {Sorel}}, \ and\ \bibinfo {author} {\bibfnamefont
  {J.}~\bibnamefont {Danckaert}},\ }\href {\doibase
  10.1103/PhysRevLett.101.093903} {\bibfield  {journal} {\bibinfo  {journal}
  {Phys. Rev. Lett.}\ }\textbf {\bibinfo {volume} {101}},\ \bibinfo {pages}
  {093903} (\bibinfo {year} {2008})}\BibitemShut {NoStop}%
\bibitem [{\citenamefont {der Sande}\ \emph {et~al.}(2008)\citenamefont {der
  Sande}, \citenamefont {Gelens}, \citenamefont {Tassin}, \citenamefont
  {Scir\'e},\ and\ \citenamefont {Danckaert}}]{Sande2008}%
  \BibitemOpen
  \bibfield  {author} {\bibinfo {author} {\bibfnamefont {G.~V.}\ \bibnamefont
  {der Sande}}, \bibinfo {author} {\bibfnamefont {L.}~\bibnamefont {Gelens}},
  \bibinfo {author} {\bibfnamefont {P.}~\bibnamefont {Tassin}}, \bibinfo
  {author} {\bibfnamefont {A.}~\bibnamefont {Scir\'e}}, \ and\ \bibinfo {author}
  {\bibfnamefont {J.}~\bibnamefont {Danckaert}},\ }\href
  {http://stacks.iop.org/0953-4075/41/i=9/a=095402} {\bibfield  {journal}
  {\bibinfo  {journal} {Journal of Physics B: Atomic, Molecular and Optical
  Physics}\ }\textbf {\bibinfo {volume} {41}},\ \bibinfo {pages} {095402}
  (\bibinfo {year} {2008})}\BibitemShut {NoStop}%
\bibitem [{\citenamefont {Heinonen}, \citenamefont {Zhou},\ and\ \citenamefont
  {Li}(2013)}]{Heinonen2013b}%
  \BibitemOpen
  \bibfield  {author} {\bibinfo {author} {\bibfnamefont {O.}~\bibnamefont
  {Heinonen}}, \bibinfo {author} {\bibfnamefont {Y.}~\bibnamefont {Zhou}}, \
  and\ \bibinfo {author} {\bibfnamefont {D.}~\bibnamefont {Li}},\ }\href@noop
  {} {\bibfield  {journal} {\bibinfo  {journal} {arXiv:1310.6791}\ } (\bibinfo
  {year} {2013})}\BibitemShut {NoStop}%
\bibitem [{Sup()}]{SuppMat}%
  \BibitemOpen
  \href@noop {} {\bibinfo  {journal} {See Supplemental material XXXXX for
  additional information}\ }\BibitemShut {NoStop}%
\bibitem [{\citenamefont {Ott}(1994)}]{Ott1994}%
  \BibitemOpen
\bibfield  {journal} {  }\bibfield  {author} {\bibinfo {author} {\bibfnamefont
  {E.}~\bibnamefont {Ott}},\ }\href@noop {} {\emph {\bibinfo {title} {Chaos in
  dynamical systems}}}\ (\bibinfo  {publisher} {Cambridge University Press},\
  \bibinfo {year} {1994})\BibitemShut {NoStop}%
\bibitem [{\citenamefont {Silva}\ and\ \citenamefont
  {Keller}(2010)}]{Silva2010}%
  \BibitemOpen
  \bibfield  {author} {\bibinfo {author} {\bibfnamefont {T.}~\bibnamefont
  {Silva}}\ and\ \bibinfo {author} {\bibfnamefont {M.}~\bibnamefont {Keller}},\
  }\href {\doibase 10.1109/TMAG.2010.2044583} {\bibfield  {journal} {\bibinfo
  {journal} {Magnetics, IEEE Transactions on}\ }\textbf {\bibinfo {volume}
  {46}},\ \bibinfo {pages} {3555 } (\bibinfo {year} {2010})}\BibitemShut
  {NoStop}%
\bibitem [{\citenamefont {Keller}\ \emph {et~al.}(2010)\citenamefont {Keller},
  \citenamefont {Pufall}, \citenamefont {Rippard},\ and\ \citenamefont
  {Silva}}]{Keller2010}%
  \BibitemOpen
  \bibfield  {author} {\bibinfo {author} {\bibfnamefont {M.~W.}\ \bibnamefont
  {Keller}}, \bibinfo {author} {\bibfnamefont {M.~R.}\ \bibnamefont {Pufall}},
  \bibinfo {author} {\bibfnamefont {W.~H.}\ \bibnamefont {Rippard}}, \ and\
  \bibinfo {author} {\bibfnamefont {T.~J.}\ \bibnamefont {Silva}},\ }\href
  {\doibase 10.1103/PhysRevB.82.054416} {\bibfield  {journal} {\bibinfo
  {journal} {Phys. Rev. B}\ }\textbf {\bibinfo {volume} {82}},\ \bibinfo
  {pages} {054416} (\bibinfo {year} {2010})}\BibitemShut {NoStop}%
\bibitem [{\citenamefont {Liu}\ \emph {et~al.}(2001)\citenamefont {Liu},
  \citenamefont {Lin}, \citenamefont {Huang}, \citenamefont {Guo},\ and\
  \citenamefont {Duan}}]{Liu2001}%
  \BibitemOpen
  \bibfield  {author} {\bibinfo {author} {\bibfnamefont {Y.}~\bibnamefont
  {Liu}}, \bibinfo {author} {\bibfnamefont {J.}~\bibnamefont {Lin}}, \bibinfo
  {author} {\bibfnamefont {G.}~\bibnamefont {Huang}}, \bibinfo {author}
  {\bibfnamefont {Y.}~\bibnamefont {Guo}}, \ and\ \bibinfo {author}
  {\bibfnamefont {C.}~\bibnamefont {Duan}},\ }\href
  {http://www.opticsinfobase.org/abstract.cfm?URI=JOSAB-18-5-666} {\bibfield
  {journal} {\bibinfo  {journal} {Journal of the Optical Society of America B}\
  }\textbf {\bibinfo {volume} {18}},\ \bibinfo {pages} {666} (\bibinfo {year}
  {2001})}\BibitemShut {NoStop}%
\bibitem [{\citenamefont {Banerjee}, \citenamefont {Bhattacharya},\ and\
  \citenamefont {Chakrabarti}(2000)}]{Banerjee2000}%
  \BibitemOpen
  \bibfield  {author} {\bibinfo {author} {\bibfnamefont {S.}~\bibnamefont
  {Banerjee}}, \bibinfo {author} {\bibfnamefont {R.}~\bibnamefont
  {Bhattacharya}}, \ and\ \bibinfo {author} {\bibfnamefont {C.}~\bibnamefont
  {Chakrabarti}},\ }\href
  {http://www.hindawi.com/journals/ijmms/2000/657092/cta/} {\bibfield
  {journal} {\bibinfo  {journal} {International Journal of Mathematics and
  Mathematical Sciences}\ }\textbf {\bibinfo {volume} {23}},\ \bibinfo {pages}
  {435 } (\bibinfo {year} {2000})}\BibitemShut {NoStop}%
\end{thebibliography}
%

\end{document}